# The Role of Cell-Matrix Interactions in Connective Tissue Mechanics

Iain Muntz[1]*, Michele Fenu[2]*, Gerjo J. V. M. van Osch[2]**, Gijsje H. Koenderink[1]**

[1] Department of Bionanoscience, Kavli Institute of Nanoscience Delft, Delft University of Technology, Van der Maasweg 9, Delft, the Netherlands

[2] Department of Orthopaedics and Sports Medicine, Erasmus MC, University Medical Center Rotterdam, Rotterdam, the Netherlands

* These authors contributed equally to this work

** Corresponding authors:  Gijsje H. Koenderink, g.h.koenderink@tudelft.nl

Gerjo J. V. M van Osch, g.vanosch@erasmusmc.nl

**Abstract:** *Living tissue is able to withstand large stresses in everyday life, yet it also actively adapts to dynamic loads. This remarkable mechanical behaviour emerges from the interplay between living cells and their non-living extracellular environment. Here we review recent insights into the biophysical mechanisms involved in the reciprocal interplay between cells and the extracellular matrix and how this interplay determines tissue mechanics, with a focus on connective tissues. We first describe the roles of the main macromolecular components of the extracellular matrix in regards to tissue mechanics. We then proceed to highlight the main routes* via *which cells sense and respond to their biochemical and mechanical extracellular environment. Next we introduce the three main routes* via *which cells can modify their extracellular environment: exertion of contractile forces, secretion and deposition of matrix components, and matrix degradation. Finally we discuss how recent insights in the mechanobiology of cell-matrix interactions are furthering our understanding of the pathophysiology of connective tissue diseases and cancer, and facilitating the design of novel strategies for tissue engineering.*

1. Introduction

Our body is literally held together by connective tissues, which support and connect our various organs and body parts. Examples are the dermal tissue that covers us, bony skeleton that supports us, cartilage that protects our joints, and tendons that connect muscles to bone. Connective tissues are composed of a non-living scaffold of protein and polysaccharide polymers known as the extracellular matrix, which surrounds and nurtures the cells that live inside. Individually, the extracellular matrix and the cells already have remarkable physical properties, but it is only when they are brought together that the full physical properties of tissues emerge. One remarkable trait of connective tissues is their high mechanical strength, which is essential to preserve their integrity in the face of constant mechanical loading. For example, while walking, muscles, tendons and skin are subject to large stretching forces while cartilage is compressed [1]. At the same time, however, tissues also need to maintain a capacity for dynamic rearrangements and growth. This contradictory combination of strength and dynamic adaptability is unique to living matter and has formed the inspiration for biophysical research into the underlying physical mechanisms.

Tissues mainly derive their mechanical strength from the extracellular matrix, a composite material consisting of filamentous proteins, which provide a mechanical scaffold, and glycosaminoglycans, which confer an adaptability to compression and allow the protein fibres to move relative to one another. There is growing evidence that synergistic effects that emerge from the combination of these two key structural components are necessary to provide tissue-specific mechanical properties required for proper function [2]. Furthermore, the unique capacity of tissues for dynamic rearrangements and growth derives from the presence of cells embedded within the extracellular matrix. Cells directly adhere to the matrix *via* transmembrane receptors that enable cells to sense and respond to the matrix but also to manipulate the matrix. Cells constantly probe the viscoelastic response of the matrix by exerting contractile forces (known as traction forces) and transduce this information into cellular decisions. Simultaneously, the traction forces enable the cells to physically rearrange or apply prestress to the matrix, changing its architecture and rigidity [3]. Additionally, cells biochemically remodel the matrix by selectively synthesising but also enzymatically degrading matrix constituents. Through a dynamic combination of mechanical forces and biochemical remodelling, cells thus adapt the extracellular environment in healthy tissue while maintaining tissue properties that are tailored to perform the required function [4]. For example bone, despite its deceptively rigid and inert appearance, is a highly dynamic tissue capable of mechanoadaptation, as shown by loss of bone mass in space and re-gain of bone mass upon physical exercise [5].

Understanding the biophysical processes that govern tissue mechanics and cell-matrix interactions is interesting in its own right, but it is also directly relevant for understanding the pathophysiology of diseases affecting connective tissues as well as for tissue engineering aimed at tissue repair or organ-on-chip disease models. It is increasingly appreciated that diseases such as fibrosis, osteoarthritis, and cancer involve malfunctional tissue mechanics originating from misregulation of cell-matrix interactions. In order to develop technologies for early diagnosis and effective treatments, it is essential to identify the mechanisms that deregulate the normal mechanochemical cell-matrix interplay. Similarly, there has been a growing appreciation that engineering and repair of tissues should make sure to use materials that mimic the complex mechanical properties of the target tissue.

In this review we will describe the different mechanisms by which cells and the extracellular matrix interact and together determine the mechanical behaviour of connective tissues. We first discuss the unique mechanical properties of tissues, focussing on universal features that are inherent to the polymeric nature of the extracellular matrix as well as molecular mechanisms that customise tissue mechanics to different functions. We then describe the different pathways *via* which cells sense and respond to their surroundings. Next we explain how cells remodel the extracellular environment via mechanical and biochemical mechanisms. Finally, we highlight the relevance of this knowledge in understanding pathophysiological mechanisms and in designing effective tissue engineering approaches. We conclude with a brief outlook of the future directions for research into connective tissue biophysics.

## 2. Matrix Mechanics

Although different connective tissues are composed of a similar core set of fibrous proteins and glycosaminoglycans, the mechanics may differ greatly depending on the specific make-up of the tissue. For example, bone tissue is characterised by the presence of stiff hydroxyapatite crystals embedded in the extracellular matrix, which confer to this tissue its strength, rigidity and brittleness [5]. On the other hand, cartilage contains a large fraction of glycosaminoglycans, conferring completely different mechanical properties such as a remarkable adaptability to compressive forces [6]. Besides differences in tissue make-up, the architecture of the constituents plays a key role. For example, collagen fibrils can be aligned, leading to the high tensile strength of tendon [7], disordered, conferring high tear resistance to skin for a wide range of loading directions [8], or organised in arcs, aiding the uniform transmission of loads to the underlying bone in articular cartilage [9]. In this section we will briefly introduce the main components of the extracellular matrix and how these affect the mechanical behaviour of connective tissues.

The principal structural components of connective tissues are fibrous proteins, such as collagen, elastin and fibronectin [10]. These proteins form a space-spanning filamentous network with different mechanical properties depending on the network composition and architecture. Collagen, the most abundant protein in vertebrate tissue [11], is produced by cells as a long (300 nm) and thin (1.5 nm) triple helical monomer that self-assembles into fibres in an axially staggered manner (Figure 1a). The so-called D-periodicity arising from the axial staggering is key for fibre strength and encodes cell recognition and binding sites for other extracellular matrix components [12,13]. Elastin is an elastomeric protein that is responsible for the high extensibility and recoil of tissues such as skin and arteries [8,14]. Elastin is present in the form of elastic fibres that are composed of approximately 90% elastin and otherwise primarily fibrillar glycoproteins (Figure 1b). Elastin's subunit, tropoelastin, has alternating hydrophobic domains, which drive self-assembly and give elasticity and recoil, and hydrophilic domains, which are responsible for cross-linking [15]. Finally, fibronectin forms flexible fibres that are extremely extensible [16]. Fibronectin polymerisation is a cell-dependent process that requires direct interactions with integrin receptors (Figure 1c) [17]. Cell-generated traction forces stretch the fibronectin fibres and expose cryptic binding sites that in turn regulate cell adhesion, migration, growth and differentiation [18]. Collagen and fibronectin assembly are thought to be interdependent processes, whereby relaxed fibronectin fibrils act as templates for collagen fibril assembly while collagen fibres shield fibronectin fibres from being stretched by cellular traction force [19,20].

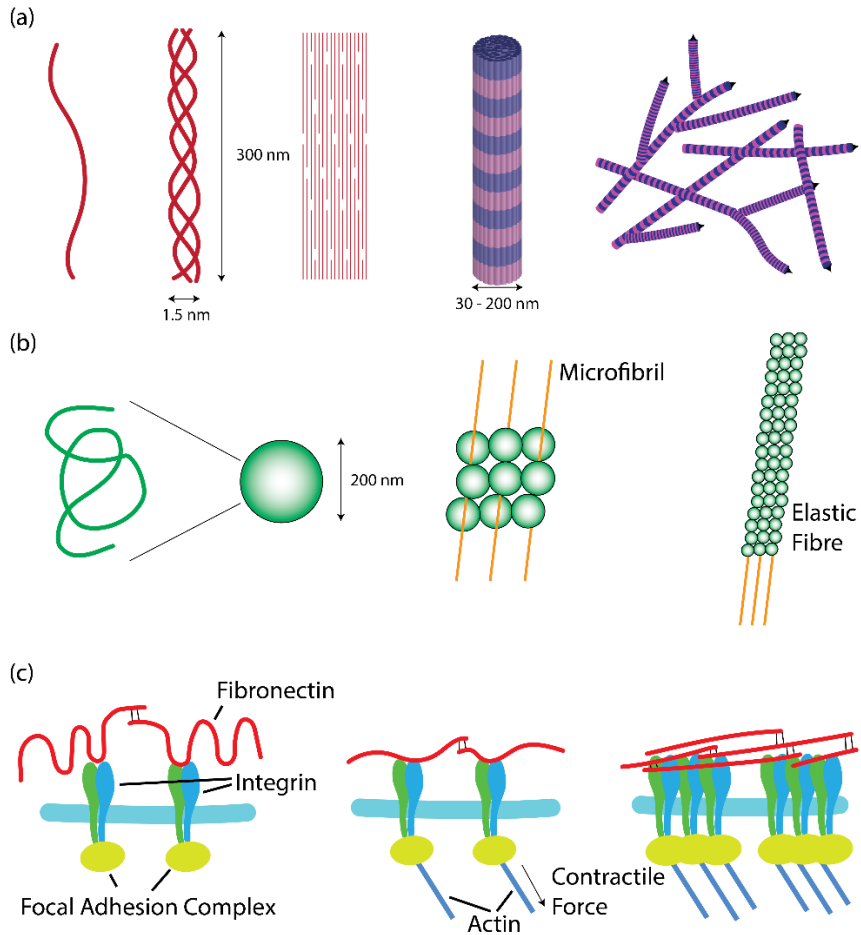

*Figure 1: The assembly of fibrous proteins is a multi-step process leading to hierarchical structures. Each image shows a schematic representation of key steps in protein assembly starting from monomeric forms (left) to the final network (right) (a) Assembly of collagen: three collagen polypeptides form a triple-helical collagen monomer intracellularly. Collagen monomers associate in an axially staggered manner extracellularly, forming thick rope-like fibres. The final fibre can then cross-link with other fibres to form a space spanning network. Note that this schematic represents collagen assembly in solution, but the assembly in vivo involves cellular processes. (b) Assembly of elastin: elastin forms thin, coiled monomeric polymer chains that undergo phase separation (coacervation) to form spherical aggregates. These spherules are deposited onto a scaffold of fibrillin, a protein found in connective tissue. Eventually elastic fibres are formed which can then cross-link to form a cross-linked network. The assembly of elastin into elastic fibres is also affected by cellular processes. (c) Assembly of fibronectin at the cell surface. Fibronectin attaches to integrins in the protein's inactive state. Contractile forces from the actin cortex activate the fibronectin. Many integrins cluster at the membrane, promoting fibronectin-fibronectin interactions leading to fibril assembly.*

Networks of fibrous proteins tend to exhibit a highly nonlinear mechanical response. Disordered collagen networks as found in skin, for example, exhibit a strain stiffening response resulting from the specific network architecture (Figure 2a). This architecture is characterised by stiff fibres with an average connectivity, $Z$, ranging from 3 (corresponding to branches) to 4 (corresponding to a pair of crosslinked fibres). While this is below the Maxwell criterion for mechanical stability of spring networks ($Z$=6 in 3D), the large bending rigidity of the fibres stabilises the network [21,22]. As a consequence, disordered collagen networks have a low elastic modulus at low strain governed by fibre bending and reorientation along the direction of strain, and undergo a transition to a rigid state at high strain governed by fibre stretching (Figure 2a). Aligned collagen networks as found in tendon also strain-stiffen, but in this case the soft response at low strain is due to initial fibre straightening [23]. In case of elastomers like elastin and fibronectin, the nonlinear mechanical response also involves conformational changes of the protein at high extensions [15,16]. Fibrin, a protein that forms a temporary extracellular matrix that promotes wound healing upon vessel injuries, has a very unusual nonlinear response that is due to a combination of its rigid fibre architecture and strain-induced conformational transitions. This behaviour arises as fibrin monomers contain both structured domains, which unfold at large tensile strains, and unstructured regions, which are flexible and can easily stretch [24,25]. The hierarchical network architecture of fibrin thus confers a multi-phase strain stiffening behaviour with increasing strain [26]. Irrespective of the underlying mechanism, the strain-stiffening response of the extracellular matrix strongly affects cell-matrix interactions as cells exert traction forces that lead to gradients in matrix stiffness around a cell [3].

The mechanical properties of fibrous networks are not only dependent on the amplitude of the deformation, but also on the time scale. There are two major mechanisms that introduce time-dependency, namely viscoelasticity and poroelasticity. Viscoelastic effects arise when the crosslinks holding the networks together are transient so relaxation of the network can occur. This is often the case in reconstituted networks such as collagen and fibrin [27,28]. Viscoelastic effects not only make the mechanical response time-dependent, but also cause the mechanical response to change with repeated large-strain loading, and influence cell-mediated matrix remodelling (Figure 2b) [29,30]. In adult tissues, viscoelastic relaxation is limited because the collagen network is chemically crosslinked by enzymes and by glycation end products that accumulate with age and chronic diseases [31]. Poroelastic effects have to do with the presence of the background fluid. When fibrous networks are subject to deformations that change the volume (i.e., tensile or compressive deformations), the incompressibility of the interstitial fluid will provoke a fluid flow through the network. This causes a time-dependent mechanical response that is referred to as poroelasticity (Figure 2c) [32]. Under fast deformations, the system will respond as an incompressible material, whereas under slow deformations, where the fluid has time to flow in or out, the system responds like a compressible material. The characteristic time scale $\tau$ for a fluid of viscosity $\eta$ to flow across a distance $d$ through a polymer network with pore size $\xi$ and shear modulus $G$ is given by

$$\tau \approx \eta d^2/Gk, \qquad [1]$$

where $k \sim \xi^2$ is the network's hydraulic permeability [33]. Poroelastic effects are particularly important and well-studied in cartilage and the intervertebral disc [34]. Even under a volume-conserving shear deformation, poroelasticity contributes to the mechanical response. Sheared polymer networks develop a normal force perpendicular to the direction of shear, which tends to be negative (contractile) for the rigid biopolymers found in the extracellular matrix [35]. Due to the presence of the fluid, this contractile effect is only seen at times that are long enough for the fluid to flow [36]. By contrast, at short times $t < \tau$, the normal stress is positive because of the strong viscous coupling between the polymer network and the interstitial fluid.

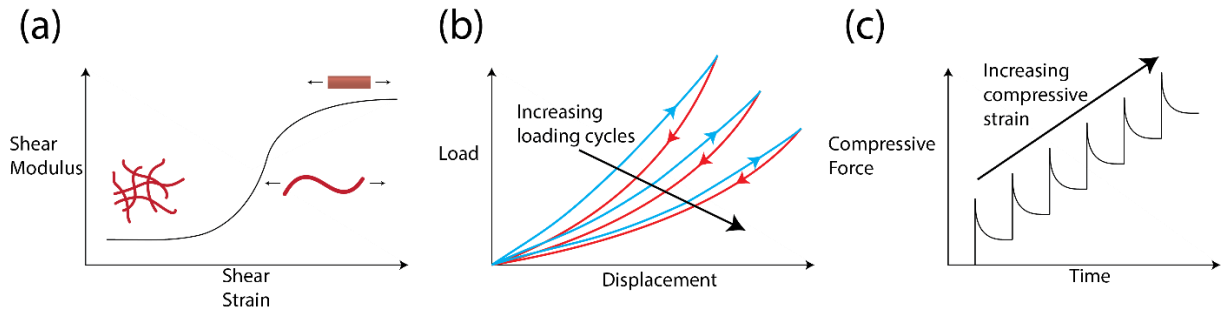

*Figure 2: The complex mechanical properties of tissues. (a) Strain Stiffening. Networks of fibrous proteins, key mechanical components of tissues, often become stiffer at higher strain. This stiffening occurs due to the hierarchical nature of the assembly process which leads to different stretching modes of the network, from deformation of the entire network, to straightening of individual fibres and stretching of individual fibres. (b) Viscoelasticity. Tissues are viscoelastic materials which can be characterised by their behaviour under cyclic deformation. On loading (blue lines) a larger load is imposed than on unloading (red lines). The energy dissipated in each cycle can be found as the area between the loading and unloading curves. Upon repeated loading the required load to reach the same displacement decreases as the network undergoes plastic deformations. (c) Poroelasticity. By compressing a tissue to a certain strain a stress is required to deform the network, however due to fluid flow through the porous network this stress relaxes over time. While the stress required to reach a certain strain indicates the compressive strength due to the network, the time scale of the relaxation reveals information about the network's permeability. This behaviour persists as the compressive strain is linearly ramped up.*

Besides structural proteins, an abundant component of the extracellular matrix of connective tissues are the glycosaminoglycans, which are linear polysaccharide chains made up of repeating disaccharide building blocks. Hyaluronic acid is a particularly important glycosaminoglycan because of its high molecular weight (in the megadalton range) and its ability to maintain hydration [37]. Often glycosaminoglycans are anchored to proteins, forming proteoglycans. Proteoglycans and glycosaminoglycans modulate connective tissue mechanics. For example, in cartilage the fixed negative charges of glycosaminoglycans induce an osmotic swelling pressure which, in combination with the tensile strength of collagen, confers on cartilage its compressive resistance [6]. Glycosaminoglycans also allow collagen fibrils to slide relative to one another during tensile loading in tendons, thus preventing damaging strains [38]. Additionally, glycosaminoglycans strongly modulate the poroelastic response of tissues to compressive loading as densely packed glycosaminoglycan chains will resist fluid flow in tissues such as cartilage due to their high osmotic pressure, increasing the tissue's resistance to compression [41].

### 3. How do cells sense matrix mechanics?

Cellular mechanosensing is a dynamic process where cells constantly probe the surrounding extracellular matrix *via* different receptors that act as mechanical links between the extracellular matrix and the cell. The receptors transmit mechanical information to the cytoskeleton, an intracellular protein network comprising three types of protein filaments: actin, microtubules and intermediate filaments [40]. The actin network is traditionally thought to be the primary component for cellular mechanosensing, but there is increasing evidence that the interplay between all three components is important (for in-depth reviews, see: [40–42]). For instance, while focal adhesions are connected to the actin network within the cell, microtubules play a role in promoting focal adhesion turnover [43,44]. The cytoskeleton transmits external and cell-generated forces all the way to the chromatin within the nucleus via transmembrane proteins embedded in the nuclear membrane, thus influencing gene expression [45]. In addition, recent experiments showed that also the nucleus itself is mechanosensitive [46].

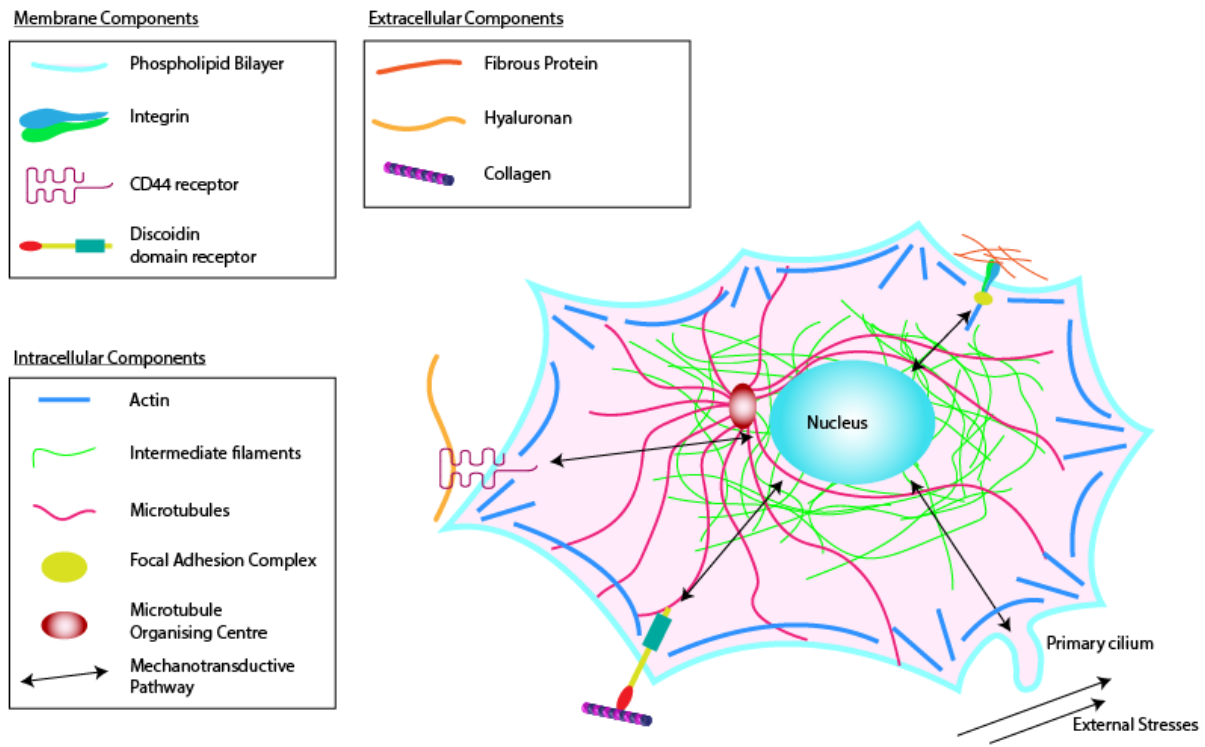

*Figure 3* Schematic overview of the main components involved in cell-matrix interactions, mechanosensing, and mechanotransduction. Cells adhere to different components of the extracellular matrix through transmembrane receptors, and these receptors are linked inside the cell to different components of the cytoskeleton. The cytoskeleton in turn is connected to the cell nucleus, thus forming a mechanical continuum. Extracellular and intracellular forces thus impact gene transcription and translation, which in turn feeds back on the cytoskeleton and matrix receptors.

Focal adhesions are complex protein-containing structures, which link the cytoskeleton (actin and, in some cases, intermediate filaments) to the extracellular matrix through transmembrane proteins called integrins. Integrins are heterodimeric, being composed of a larger $\alpha$ subunit and a smaller $\beta$ subunit. In mammalian cells there are 24 distinct $\alpha\beta$ integrin pairs which each bind to specific extracellular matrix components, for example, the $\alpha 1\beta 1$ integrin binds to collagen and laminin while the $\alpha 5\beta 1$ binds to fibronectin [47]. However, multiple integrin pairs can bind to the same extracellular components and many integrin pairs bind multiple extracellular components. The extracellular sensing of integrins is activated *via* recruitment of the protein talin to the intracellular domain of the integrin molecule [48]. When a cell adheres to a substrate, the integrin adhesion complexes mature into a focal adhesion and talin undergoes a conformational change that causes the exposure of binding sites for intracellular proteins such as actin and vinculin. On flat, rigid substrates, the adhesion complexes form large and long-lived mechanical bridges termed focal adhesions. Depending on the stiffness and geometry of the substrate, alternative integrin adhesion complexes can form such as fibrillar adhesions. These adhesions allow cells to sense the environment's physical properties such as compliance, dimensionality or ligand spacing [49]. The interpretation of these mechanical signals by the cell can lead to differing cell behaviours (mechanotransduction), e.g., in cell differentiation [50], cell proliferation [51], or cell migration [52]. Key components of intracellular signalling pathways are the YAP and TAZ proteins, a pair of homologous transcriptional coactivators [53]. The YAP/TAZ proteins are sensitive to the level of filamentous actin, as an indirect readout for mechanical force. Moreover, it was recently shown that exposure of cells to a stiff environment promotes a mechanical connection between the nucleus and the cytoskeleton, allowing forces exerted through focal adhesions to stretch nuclear pores and increase YAP nuclear import [54]. For more details on the signalling pathways involved in mechanotransduction, such as the YAP/TAZ or Rho GTPase pathways, we refer to other reviews [4].

In addition to integrin-based adhesions, cells can sense their environment through receptors specific to certain extracellular matrix components. For example, collagen is specifically sensed through discoidin domain receptors, which are essential for controlling the density and alignment of fibrillar collagen [55,56]. These receptors facilitate cellular forces on collagen fibres because they interact intracellularly with cytoskeletal myosin filaments, a motor protein which generates traction forces [56]. Hyaluronan is sensed *via* multiple receptors, the best characterised of which is CD44, a hyaladherin that mediates many cellular functions such as motility, inflammation and growth [57]. CD44 receptors promote cell adhesion and migration in hyaluronan-rich matrices via the formation of microtentacles protruding from the cells [58]. Although hyaluronan is the main ligand for CD44, this receptor can also bind to other extracellular components such as collagen or fibronectin [59]. Intracellularly, CD44 is connected to the actin cytoskeleton via the family of ezrin/radixin/moesin proteins, suggesting that CD44 is closely involved in mechanosensing [60]. Indeed, CD44 was identified as an upstream mediator of the activity of ERK, AKT and YAP pathways in cancer cells [61].

Mechanosensitive ion channels also contribute to cellular mechanotransduction [62,63]. For instance, upon mechanical stimulation of the phospholipid membrane, such as tension or compression, these channels can change conformation, switching from a closed to an open state, therefore supporting ion flows across the membrane. Furthermore, most cells of the body have immotile organelles known as primary cilia [64], long (1-3 μm) membrane protrusions composed mainly of tubulin. Upon deflection due to external stresses, mechanical input is transferred and transduced into the cell. For instance, in cartilage development and response to dynamic loading, primary cilia have been implicated in cell differentiation, proliferation and regulation of aggrecan and glycosaminoglycan deposition [65,66].

## 4. Cell-mediated modification of the extracellular matrix

Cells employ three main (mechanosensitive) mechanisms to alter the composition, structure and mechanics of the extracellular matrix. Firstly, they can stiffen their environment by applying forces via transmembrane receptors [3,67]. Secondly, cells remodel their environment *via* the deposition and crosslinking of extracellular matrix components [68]. Finally, cells degrade the extracellular matrix using both membrane-bound matrix-degrading enzymes and proteolytic enzymes that are secreted into the extracellular space [69].

### 4.1 Cells alter their matrix through contractile forces

The cytoskeletal actin network actively generates contractile forces within the cell using the activity of non-muscle myosin II motor proteins. Cell contractility can be activated through mechanotransductive signals which are sensed by the cell *via* focal adhesions and transduced *via* intracellular pathways. Therefore, cells probe but also remodel the extracellular matrix through integrins. Cell-mediated traction forces can cause significant stiffening of the matrix due to its intrinsic strain-stiffening response [70,71]. Contractile carcinoma cells have, for instance, been shown to induce stiffness gradients in surrounding collagen, fibrin or Matrigel matrices that can span two orders of magnitude when comparing regions close to or far from a cell [3]. Computer simulations have shown that cell-mediated traction forces coupled with the unique nonlinear elasticity of fibrous extracellular matrices should facilitate the transmission of forces at long distances from the cells [72]. Namely, cells establish aligned fibre tracts between each other and generate elastic anisotropies in their local environment [73]. Experimentally these fibre tracts have indeed been confirmed by time-lapse imaging [74], and there is intriguing evidence that cells can sense the orientation and position of neighbouring cells from tension-induced enhancements of the matrix stiffness [75,76] .

### 4.2 Cells alter their environment through deposition of extracellular matrix components

Cells are essential in the maintenance and mechanoadaptation of tissue integrity throughout the life of an individual [4]. They secrete extracellular matrix components that assemble in the extracellular space in a tightly regulated manner, both during tissue development and in adult tissue maintenance and remodelling [68]. In mammals, more than 300 different proteins can be found in the extracellular matrix, and more than 700 proteins have been identified to be involved in the process of matrix generation and remodelling, customizing the matrix for different tissue-specific functions [77]. For example, procollagen is pre-assembled within the cell, but the intervention of extracellular enzymes is needed to form mature collagen fibres and the diameter of these fibres is tailored to the specific requirements of each tissue by auxiliary molecules such as glycosaminoglycans [78]. In stiff tissues like tendon, the diameter of collagen fibres is around 200 nm to ensure tensile strength, whereas in the cornea, the diameter is only 30 nm to ensure optical transparency [79,80]. The complete picture of multiscale extracellular matrix assembly is yet to be elucidated [81].

Recent studies have indicated that cells deposit extracellular components in the pericellular space as early as four hours after encapsulation in a synthetic hydrogel, and the deposited layer thickens over time (see Fig. 4a; [82,83]). While this general behaviour is maintained across hydrogels with different mechanical characteristics, the thickness of the deposited layer depends on the physical properties of the hydrogel, particularly the crosslinking density [84]. Interestingly, the crosstalk between the deposited layer of nascent proteins and the cells appears to affect cell differentiation. For example, when cell-matrix interactions are inhibited, osteogenesis decreases while adipogenesis is enhanced [82,84]. A recent study characterised the mechanics of the nascent extracellular matrix deposited by chondrocytes with correlative atomic force/fluorescence microscopy. This revealed that chondrocytes deposit a pericellular matrix with similar elastic properties independently of the properties of the hydrogels [83]. This result indicates that cells tend to remodel their surroundings to recreate an optimal mechanical environment.

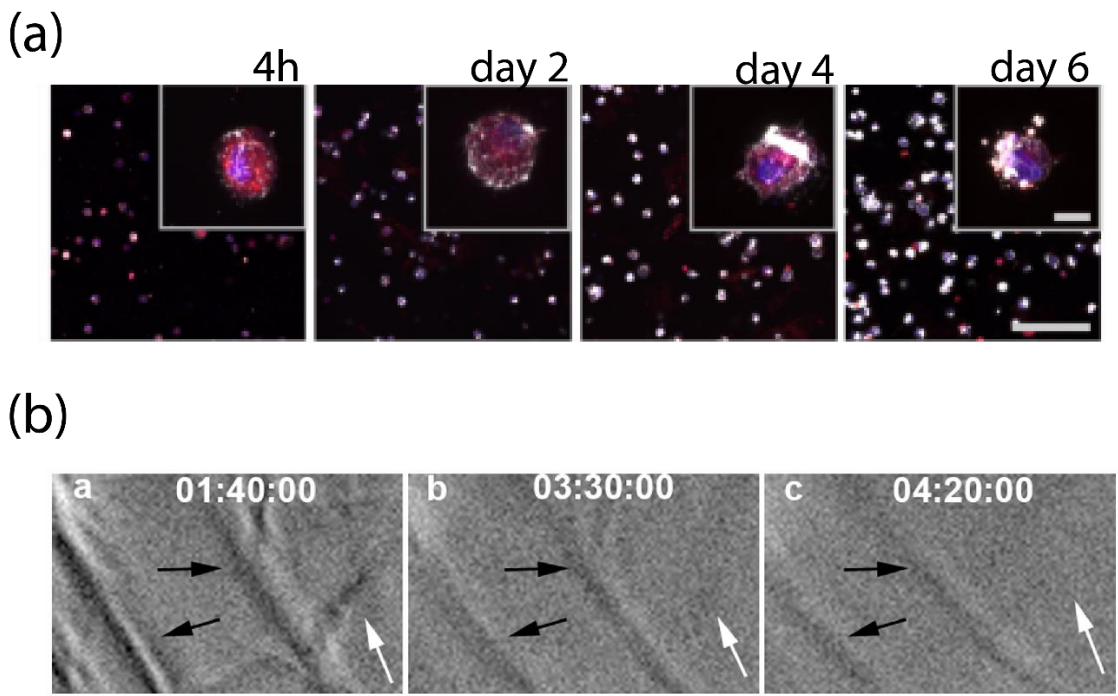

*Figure 4. Biochemical remodelling of the extracellular matrix by cells. (a) Representative images of matrix deposited by cells, which are labelled with a plasma membrane stain (red) and with DAPI for the cell nucleus (blue). Cells were encapsulated in non-degradable hyaluronan-based*

*hydrogels and cultured in growth media supplemented with an azide-modified methionine. The bio-orthogonal strain-promoted cyclo-addition between the azide and a fluorophore (DBCO-488) enables visualisation of all nascent extracellular matrix proteins. The cells already start to produce a significant amount of material within 4 hours, and continue to produce more material during the entire culture period of 6 days. Scale bars, 200 µm (main) and 20 µm (insets)* [82]. ***(b)** Tension-sensitive collagen degradation. Strained collagen networks were exposed to active matrix metalloproteinase 8 and sequentially imaged by differential interference contrast imaging. The black arrows indicate fibrils under tension while the white arrows denote unstrained fibrils. Unstrained fibrils exhibit faster degradation when compared with strained fibrils. Times are expressed as hh:mm:ss. Scale bar 5 µm* [85].

### 4.3 Cells alter the extracellular environment through secretion of matrix degrading enzymes

Besides 'constructive' remodelling through matrix deposition and collagen alignment, cells also remodel the extracellular matrix through the degradation of specific macromolecular components with proteolytic enzymes [68]. A prominent class of enzymes are the metalloproteinases, whose main families are matrix metalloproteinases (MMPs) and a disintegrin and metalloproteinase with thrombospondin motifs (ADAMTs). These families contain respectively 23 and 19 distinct enzymes which share a common structure but have a distinct catalytic domain that, in turn, determines the binding affinity and therefore the target substrate for proteolytic activity of such enzymes. For instance, MMP1 preferentially cleaves fibrillar collagen while MMP2 has a high gelatinase activity [86]. Additionally, these enzymes can show proteolytic activity against several other components of the extracellular matrix. While most of these enzymes are secreted in the extracellular environment, some, such as MMP14, are anchored to the cell membrane and are specialised in cleavage of the extracellular matrix in the pericellular space [86,87]. Several other families of enzymes have been identified. Serine proteases such as plasmin or cathepsin G cleave peptide bonds, while hyaluronidases (hyaluronidases 1-4, PH20 and HYALP1) catalyse the degradation of hyaluronic acid [88,89].

Interestingly, it was recently shown that the enzymatic degradation of fibrillar proteins such as collagen and fibrin is tension-sensitive (Figure 4b) [85,90,91]. Enzymatic cleavage is inhibited when the fibres are under tension, suggesting a mechanosensitive control mechanism encoded in the matrix by which cells can selectively alter their surrounding matrix. This conceptually interesting use-it-or-lose-it mechanism ensures that fibres necessary for bearing load remain whereas unnecessary fibres are degraded. The exact mechanisms of strain-sensitive degradation are not yet understood because of the complex hierarchical structure of the extracellular matrix. At the network level, mechanical deformation may hamper proteolytic degradation by slowing enzyme diffusivity [92], but at the single-molecule level, mechanical strain may enhance enzymatic cleavage [93]. At the fibre level, theoretical modelling suggests that, in the case of collagen, the cleavage sites may become less accessible for cell-secreted enzymes as the molecular packing structure of the fibres rearranges under tension [94]. It will be an interesting challenge to disentangle the contributions of these different levels of structural hierarchy in encoding strain-sensitivity.

Cells have an additional control mechanism by which they can inhibit enzymatic degradation of the matrix[95], namely through the production of inhibitors such as tissue inhibitors of metalloproteinases (TIMPs) or serine proteases inhibitor (serpins) [95]. While one might intuitively expect the degradation of the extracellular matrix to be greatest nearest to the cell, the opposite may be observed. This concept was demonstrated by utilising enzyme-degradable poly-ethyleneglycol hydrogels, showing that a "reverse reaction-diffusion" profile arises as TIMPs bind to the degrading enzymes [96]. The TIMP-MMP complexes then diffuse before unbinding after a matter of minutes and they degrade the matrix only after having diffused relatively far from the cell. Therefore, cells can maintain their immediate environmental mechanics while still influencing distant extracellular environments. By inhibiting TIMPs, the authors successfully caused the degradation to be greatest near the cells, which could be applied for therapeutic purposes to increase cell migration.

## 5. Relevance of tissue physical biology in disease

In native connective tissues, cells are embedded in the extracellular matrix and exposed to physiological loading. During development and adult life, resident cells interpret the mechanical input they receive to fuel a continuous turnover of matrix components in which synthesis and degradation are in balance and adaptation of properties to changed mechanical situations is possible to a certain level. In the absence of normal mechanical stimulation, such as forced immobility, but also in excessive over-loading, this balance is compromised, resulting in loss of mechanical stability and ultimately injury or tissue rupture [97]. For instance, during the development of osteoarthritis, excessive enzymatic activity results in a degradation of glycosaminoglycans, weakening the tissue and exposing the fibrous collagen network to greater strains that may cause irreversible damage [98]. As a consequence, the cells will be exposed to a different mechanical environment and, in addition, to degradation products which often alter their behaviour and their phenotype. For instance, integrin adhesion to fibronectin fragments generated during the degeneration of the extracellular matrix promotes the production and the activity of degrading enzymes, thereby feeding a catabolic degradation loop [99].

Matrix deposition by cells is dysregulated in several pathologies such as fibrosis or cancer [100]. In fibrosis, chronic inflammation leads to stiffening of the extracellular environment because the rate of matrix deposition exceeds the rate of degradation. The unregulated matrix deposition also perturbs the normal highly organised arrangement of the matrix, thus compromising the mechanical function and associated functionality of the whole tissue [101]. In tendinopathies, for instance, the formation of scar tissue upon tissue rupture is a consequence of cells attempting to restore their mechanical environment. However, despite the cells' attempts to restore their matrix, the presence of inflammation perturbs the ratios of type I and III collagen production, leading to a stiffer tissue which is often prone to re-injury [102].

Growing knowledge on the mechanics of cell-matrix interactions has already fuelled novel hypotheses on molecular mechanisms underlying tissue degeneration. For instance, in osteoarthritis the degeneration of cartilage accompanied by a phenotypic change of chondrocytes was associated with the expression of different subsets of integrins that promote cell adhesion to fibrillar proteins [110]. Therefore, therapeutic strategies aiming at the modulation of integrin expression or the modulation of integrin binding should be investigated with the potential to stop or reverse the degenerative processes. As an aside, integrin-targeted drugs are also investigated in the context of cancer, where inhibitors of integrin binding to collagen have been shown to suppress cancer cell proliferation and migration [104,105].

## 6. Relevance of tissue physical biology in tissue engineering

The capacity of cells to secrete and remodel extracellular components has been exploited in the field of tissue engineering in attempts to engineer functional tissue *in vitro*. A landmark study in 2006 first demonstrated that the mechanical micro-environment can influence cell differentiation [103]. Meanwhile it has been shown that cells also secrete extracellular components dependent on their environment and the interplay between matrix stiffness and new matrix deposition dictates cell fate [82]. The realisation that tissue formation is mechanosensitive has led to the development of bioreactors that apply mechanical stimuli during tissue culture. For instance, in tissue engineered articular cartilage, cells are exposed to dynamic compressive loading, which promotes secretion and deposition of collagen and glycosaminoglycans [104]. However, the natural movement of the knee exposes cartilage to shear stress in addition to compressive loading. Some studies have investigated the effect of superimposing sliding motion or shear stress to dynamic compression during cartilage tissue engineering. These have revealed that the combination of mechanical stimuli affects both cell differentiation and functional matrix deposition, resulting in higher chondrogenic gene expression and stiffer mechanical properties [105,106]. In tendon tissue engineering, cells are seeded into fibrous hydrogels and exposed to uniaxial constraints bound at the extremity of the long axis. Due to the contractile properties of the cells, the hydrogels shrink and the combination with the uniaxial constraint results in the formation of highly aligned fibrous matrices parallel to the long axis, mimicking the *in vivo* situation [107]. Knowledge of integrin binding affinities is already often integrated into the design of novel materials or the modification of existing materials to enhance cell adhesion. For instance, RGD peptides (arginine-glycine-aspartate) that are recognised by most integrins [108] are added to an otherwise non-adhesive environment such as alginate, to allow cells to bind and sense the environmental mechanical properties [109], thus directing cell differentiation and tissue formation.

Most studies evaluating the effect of (bio)materials on tissue engineering and repair have focussed on matrices composed of a single macromolecular component. However, recent studies showed that composite materials can better mimic the unique mechanical resilience of tissues, for instance by combining the tensile strength of fibrillar proteins such as collagen or fibrin with the resilience/recoil of elastin or with hyaluronan's compressive resistance [2,113]. Moreover, the novel concept of early protein production indicates that cells have a short-lived interaction with the initial matrix provided after cell encapsulation. However, little is currently known regarding the interaction between cells and the nascent layer of proteins or the interaction between the nascent layer and the artificial matrix. Several methods have recently been developed that can help shed light on this issue. For instance, optical micro-rheological techniques using tracer particles seeded in the tissue have been successfully employed to study local variations of the mechanical environment in the pericellular space [114]. Furthermore, metabolic labelling techniques based on the principle of click chemistry offer the opportunity to evaluate the dynamics of secreted extracellular components, while cytoskeletal dyes compatible with live microscopy offer the chance to observe cytoskeletal rearrangements during prolonged tissue culture [82].

## 7. Conclusion

In this review we have provided an overview of recent literature concerning cell-matrix interactions and their role in connective tissue mechanics. Connective tissues provide a rich source of inspiration for physicists, presenting poorly explored questions in polymer physics, active soft matter physics, and the interplay of physics with (bio)chemistry or (mechano)biology. At the same time, biophysical research on tissues and cell-matrix interactions can provide mechanistic insights that benefit other fields. In the past years we have seen a growing interest in understanding not only the mechanical behaviour of the extracellular matrix, but also force transmission and exertion between cells and the extracellular environment and how cells can alter the composition, structure and topographical arrangement of their surrounding environment. The continued development and application of new techniques to unravel cell-matrix interactions is key to provide necessary understanding of the function and pathophysiology of both native and engineered tissues and to help develop improved treatments.

**Glossary**

Extracellular matrix: The polymeric network of fibrous proteins, glycosaminoglycans and other components located outside the cell membrane.

Pericellular matrix: The region of the extracellular matrix in close proximity to the cell. The pericellular matrix can be mechanically and biochemically distinct from the rest of the extracellular matrix.

Cytoskeleton: The polymeric network of proteins inside the cell, such as actin and myosin, which provides structural integrity, generates forces, and transmits mechanical signals in a mechanical continuum from the cell membrane to the nucleus.

Focal adhesion: A transmembrane protein complex which attaches to the extracellular matrix outside the cell and the cytoskeleton inside the cell, key for transmitting external mechanical signals to the cell nucleus and cell-generated forces to the matrix.

Integrin: A dimeric transmembrane protein, which is composed of an $\alpha$ and $\beta$ subunit, that is a key component of focal adhesions.

CD44: A transmembrane receptor ubiquitous in mammalian cells which is implicated in cell-cell and cell-matrix interactions, in particular sensing hyaluronic acid.

MMPs: Matrix MetalloProteases are enzymes that are calcium-dependent zinc-containing endopeptidases, which can be secreted by cells to selectively degrade extracellular matrix proteins.

ADAMTS: A Disintegrin And Metalloproteinase with Thrombospondin Motifs, enzymes which degrade certain components of the extracellular matrix

Glycosaminoglycan: Polysaccharide consisting of repeating disaccharide units. The four primary groups of glycosaminoglycan are classified based on their core disaccharide units and include heparin/heparan sulfate, chondroitin sulfate/dermatan sulfate, keratan sulfate, and hyaluronic acid. Their function in tissues is widespread: they confer mechanical properties to connective tissues such as shock absorption or compressive resistance and also play biological roles in binding growth factors and modulating different cell processes.

Primary cilium: An organelle which protrudes from a cell to facilitate direct sensing of external stresses such as fluid flows.

Mechanosensing: The ability of a cell to sense mechanical stimuli from the surrounding environment.

Mechanotransduction: The ability of a cell to convert mechanical signals to biological processes.

Osteoarthritis: A disease affecting joints characterised by a continuous degenerative process where the degradation, deposition and modification of extracellular components is compromised. This affects the cellular phenotype and one or more components of the extracellular matrix which causes the tissue to eventually lose its functionality.

Fibrosis: A disorder affecting tissues, during a reparative or reactive process, where deposition of nascent matrix components is compromised in such a way that there is excessive deposition. This leads to considerable stiffening of the tissue and hampers the tissue's proper function.

Tissue engineering: "An interdisciplinary field which applies the principles of engineering and life sciences toward the development of biological substitutes that restore, maintain, or improve tissue function" [115].

Matrix Deposition: The process by which cells secrete extracellular matrix components which then assemble in the extracellular matrix.

Matrix Degradation: The process by which cells degrade their surroundings, for instance to facilitate cell migration or to remodel the matrix to cope with changing requirements.

Traction force: Contractile forces generated by actin-myosin in the cytoskeleton applied by cells on their surroundings.

Collagen: The most abundant protein in mammals that forms fibrils with a characteristic hierarchical structure. Type I collagen is the most abundant member of the collagen family, which comprises both fibrous and nonfibrous collagens.

Elastin: An extracellular protein that allows tissues to return to their original shape even after large and repeated deformation.

Fibronectin: An extracellular protein which plays a large role in binding to cells and linking cells to other extracellular components.